# Giant reversible barocaloric effects in Nitrile Butadiene Rubber around room temperature

E. O. Usuda[1,2], W. Imamura[1,3], N. M. Bom[1], L. S. Paixão[1] and A. M. G. Carvalho[1]

[1]*Laboratório Nacional de Luz Síncrotron (LNLS), Centro Nacional de Pesquisa em Energia e Materiais (CNPEM),CEP 13083-100, Campinas, SP, Brazil*

[2]*Departamento de Ciências Exatas e da Terra, Universidade Federal de São Paulo (UNIFESP), CEP 00972-270, Diadema, SP, Brazil.*

[3]*Faculdade de Engenharia Mecânica, Universidade Estadual de Campinas (UNICAMP), CEP 13083-860, Campinas, SP, Brazil.*

## ABSTRACT

Elastomers have shown to be promising barocaloric materials, being suitable candidates for solid-state cooling devices. Moreover, this family of polymers presents additional advantages, such as their low cost and long fatigue life. In this context, we investigated the barocaloric effects in Nitrile Butadiene Rubber (NBR) in a large range around room temperature. Moderated applied pressures on NBR yield giant temperature change ($\Delta T_S$) and entropy change ($\Delta S_T$), reaching the maximum values of 16.4(2) K at 323 K and 59(6) J kg$^{-1}$K$^{-1}$ at 314 K, respectively, for a pressure change of 390 MPa. Besides, both $\Delta T_S$ and $\Delta S_T$ have shown to be rather reversible. An influence of the glass transition on the barocaloric effects was verified: the glassy state tends to diminish the entropy and temperature changes in comparison with the rubbery state. Furthermore, we calculated the pressure coefficient of glass transition ($dT_g/dP$) obtained from different processes. Our study evidences the potential of NBR for cooling applications based on barocaloric effect, but also points out the glass transition must be avoided for a better barocaloric performance.

## I. INTRODUCTION

The Nitrile Butadiene Rubber (NBR) is a well-known polymer from the family of elastomers with remarkable mechanical and chemical properties. NBR is an unsaturated copolymer formed by chains composed of butadiene ($C_4H_6$) and acrylonitrile ($C_3H_3N$). The rubber-like behavior (elasticity), the resistance to chemical agents (e.g. oils and acids) and low-cost make the NBR an interesting material for applications in many fields, such as automotive and aeronautic industry, as well as for making disposable equipment [1]. In recent years, elastomers have been found to be promising materials for applications in solid-state cooling [2–5]. Despite the favorable properties of NBR, this elastomer has not been explored in view of its cooling potential to date.

Current refrigeration technology is based in vapor-compression cycles, which brings environmental and energetic issues. Aiming to solve these problems, solid-state cooling devices appear as promising options [6–8]. This technology is based in materials which present *i*-caloric effects, in other words, materials exhibiting a thermal response when exposed to an external field change (*i* stands for intensive thermodynamic variables – denoting the external fields). The nature of the external field can be magnetic (*H*), electric (*E*) or mechanical (*σ*). Therefore, we can categorize the *i*-caloric effects as: magnetocaloric (*h-*

CE), electrocaloric ($e$-CE) and mechanocaloric effects ($\sigma$-CE). The later can be subdivided in barocaloric ($\sigma_b$-CE) and elastocaloric effect ($\sigma_e$-CE). Over the past two decades, the research on caloric materials have seen a fast growth due to the discoveries of the giant $h$-CE in Gd$_5$Si$_2$Ge$_2$ compound in 1997 [9], the giant $e$-CE in PbZr$_{0.95}$Ti$_{0.05}$O$_3$ in 2006 [10] and the giant $\sigma_b$-CE in Ni-Mn-In shape-memory alloy in 2010 [11]. However, the studies on $i$-caloric materials date from 1805, when J. Gough reported a temperature change in natural rubber under rapid stretching ($\sigma_e$-CE) [12]. Decades later, W. Thomson predicted the $\sigma$-CE, $h$-CE and $e$-CE using thermodynamic considerations [13,14].

Regarding the $\sigma_b$-CE, it is the least studied $i$-caloric effect so far. Only a small number of materials with giant $\sigma_b$-CE have been reported in the literature, such as: some shape-memory alloys (SMA) [11,15], fluorites [16–20], magnetic materials [21–23], ferri/ferroelectric materials [24,25] and a hybrid perovskite [26]. Polymers also exhibit potential as great $\sigma_b$-CE materials. In 1982, a giant $\sigma_b$-CE was measured in poly(methyl methacrylate) [27]. More recently, interesting results were reported for elastomers: vulcanized natural rubber (V-NR) [3] and polydimethylsiloxane (PDMS) [4] showed giant $\sigma_b$-CE around room temperature. Furthermore, supergiant $\sigma_b$-CE was measured in acetoxy silicone rubber [5]. One advantage of elastomers is that they exhibit giant $\sigma_b$-CE around room temperature even in absence of phase transitions. This yields to very high and wide table-like $\sigma_b$-CE [2–4].

In this context, NBR appears as a new option of barocaloric material for application. Here, we investigate the $\sigma_b$-CE in NBR in a large range around room temperature at moderate pressures (up to 390 MPa). Direct measurements of temperature change ($\Delta T_S$) yielded giant values. Strain ($\varepsilon$) versus temperature (T) data allowed us to calculate the isothermal entropy change ($\Delta S_T$). Moreover, we observed both $\Delta T_S$ and $\Delta S_T$ results are clearly hindered by the glass transition, which shifts to higher temperature with the applied pressure.

## II. EXPERIMENT

We used a commercial vulcanized NBR sample reinforced with carbon black fillers provided by Elastim company. The material was supplied in a long cylindrical shape with a diameter of 12 mm. The sample with 8 mm in diameter and 20 mm in length was formed from the supplied material. The density of the sample, measured with a pycnometer, is 1390(10) kg m$^{-3}$. We characterized the NBR sample via Fourier transform infrared spectroscopy (see Fig. 1 in the Supplemental Material [28]) using a PerkinElmer spectrometer (model spectrum Two).

The experimental setup and procedures used in the present work are described in detail elsewhere [2–4,29]. Strain vs. temperature experiments were performed through isobaric processes, i.e., the temperature was varied, in a rate of 4 K/min, between 213-333 K under constant pressures within the 4 – 390 MPa range.

Direct $\Delta T_S$ measurements were performed by applying/releasing pressure (maximum values within the range of 26 – 390 MPa) in a quasi-adiabatic condition. When the temperature in the sample is stable at the set point, a compressive stress is rapidly applied, resulting in a sharp increase in temperature. The load is kept constant until the temperature downs to the initial value. Finally, the stress is fastly released, causing an abrupt decrease in the sample's temperature.

# III. RESULTS AND DISCUSSION

$\Delta T_S$ data as a function of the initial temperature for compression and decompression processes are shown in Fig. 1(a). Both processes present very close results in the compared temperature range (293 – 324 K). At 390 MPa, we observe a maximum $\Delta T_S$ value of 16.4(2) K (at 323 K) for decompression. This giant barocaloric $\Delta T_S$ surpasses or it is comparable to the best barocaloric intermetallics reported so far [30]. Normalizing this result by the applied pressure ($|\Delta T_S/ \Delta\sigma|$) we have a huge value of 42(2) K GPa$^{-1}$. For the two lowest pressures, the $\Delta T_S$ curves present a slight ascending behavior as the temperatures increases. On the other hand, for 86 MPa and above, we observe a stronger dependence of the $\Delta T_S$ with temperature: the curves show a smooth increase of the $\Delta T_S$ values, tending to saturate at higher temperatures (plateau). This behavior can be attributed to the influence of the glass transition of the NBR, since the mobility of the chains is significantly reduced as the material changes from the rubbery state (plateau) to the glassy state (steep region and below). Therefore, $\Delta T_S$ values tend to decrease below the glass transition temperature ($T_g$). Corroborating this hypothesis, a similar behavior was reported by Bom *et al* in vulcanized natural rubber [3].

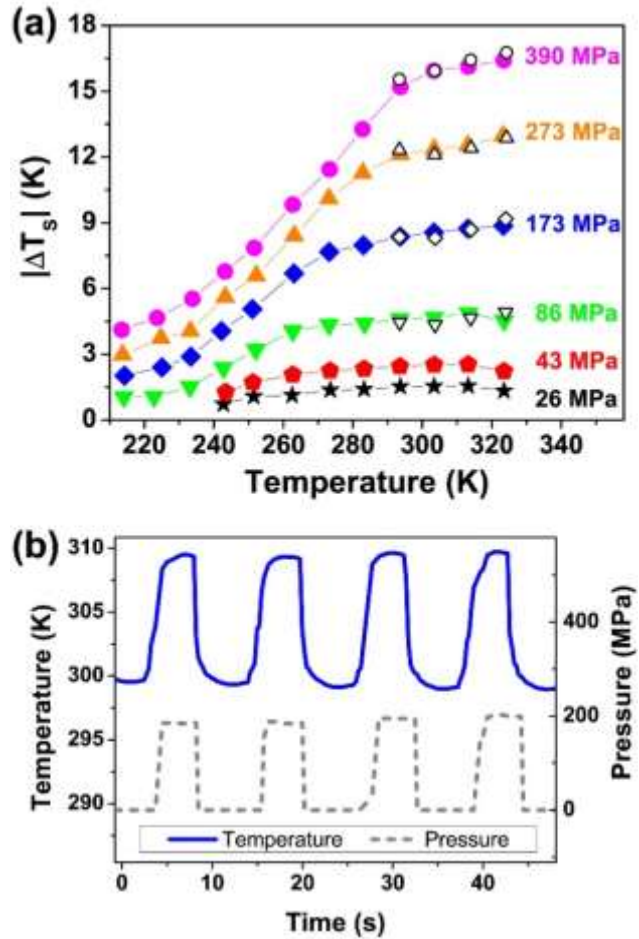

FIG. 1. (a) Adiabatic temperature change ($\Delta T_S$) vs initial temperature. Solid symbols represent the measurements for the decompression processes, while the open symbols represent the compression processes. The NBR sample was measured at different pressure variations: 26.0(5), 43(1), 86(2), 173(3), 273(8) and 390(12) MPa. The line connecting the dots are guides for the eyes. (b) Temperature (blue curve) and pressure (grey curve) vs time for four compression-decompression cycles, after several similar cycles.

In Fig. 1(b), the temperature and pressure are shown as a function of time for sequential adiabatic compression-decompression cycles. Note that the procedure is different from the measurements of $\Delta T_S$. Here, the sample does not return to the initial temperature after the pressure is applied. The behavior of the temperature change shows a rather reversible process, since the temperature change in compression is very close to the temperature change in decompression, with a difference less than 3%. Furthermore, the magnitude of the temperature change (~10 K) remains the same for repeated applications of pressure, also showing the reproducibility of these cycles.

It is a known fact that $T_g$ is sensitive to pressure [3,30–35], so it is expected that $T_g$ shifts to higher values as pressure increases. This behavior is shown in Fig. 2, where $T_g$ was calculated from $\Delta T_S$ and ε vs T. data for pressures within the range of 0 – 390 MPa. We defined $T_g$ as the point where the third derivative is zero in the curves of Fig. 1 and Strain-Temperature data (See Fig. 2 and Fig. 3 in Supplemental Material [28]). Values from strain data exhibit distinct slopes for cooling and heating processes. This discrepancy occurs due to the different kinects involving glass transformation on cooling and heating processes, which may shift the $T_g$ to higher or lower values [34,37]. In the cooling process, $T_g$ values are lower than those in the heating process, which is expected for polymers, since the material is coming from an equilibrium state (above $T_g$) to a non-equilibrium one (below $T_g$).

For $T_g$ obtained from $\Delta T_S$ data, we observe the curve is in between the cooling and heating curves. This is possibly explained by the differences in the processes [38]. On one hand, the ε vs. T was performed at isobaric process. On the other hand, $\Delta T_S$ is measured when pressure is applied very quickly increasing the sample's temperature; then, the material cools down to the initial temperature and the pressure is released. These two distinct experimental procedures may yield different results. Still, there is a good agreement between the data from both methods considering the estimated errors. The pressure coefficient of the glass transition temperature ($dT_g/dP$) for the heating, cooling and $\Delta T_S$ curves are 0.186(4), 0.14(2) and 0.11(4) K MPa$^{-1}$, respectively.

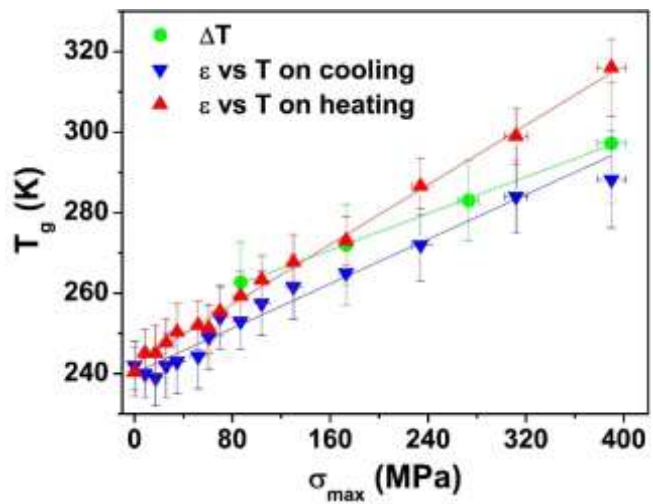

FIG. 2. Glass transition temperature ($T_g$) vs pressure (σ) for NBR. Up triangles and down triangles correspond to $T_g$ calculated from ε vs T data for heating and cooling processes, respectively. Circles are calculated from $\Delta T_S$ vs T data. The solid lines represent the linear fit for each curve.

The entropy change was indirectly obtained by taking the derivative $(d\varepsilon/dT)_\sigma$ from strain vs. T curves (See Fig. 2 and Fig. 3 in Supplemental Material [28]) and using Maxwell's relation to reach the following expression [39,40]:

$$\Delta S_T(T, \Delta\sigma) = -\frac{1}{\rho_0} \int_{\sigma_1}^{\sigma_2} \left(\frac{\partial \varepsilon}{\partial T}\right)_\sigma d\sigma,$$

where $\sigma$ and $\rho_0$ are the compressive stress and the density of the sample at atmospheric pressure and room temperature, respectively. $\varepsilon$ is the strain, defined as $\varepsilon \equiv \Delta l/l_0$, where $\Delta l$ is the length change of the sample and $l_0$ is the initial length, a constant measured at ambient pressure and room temperature.

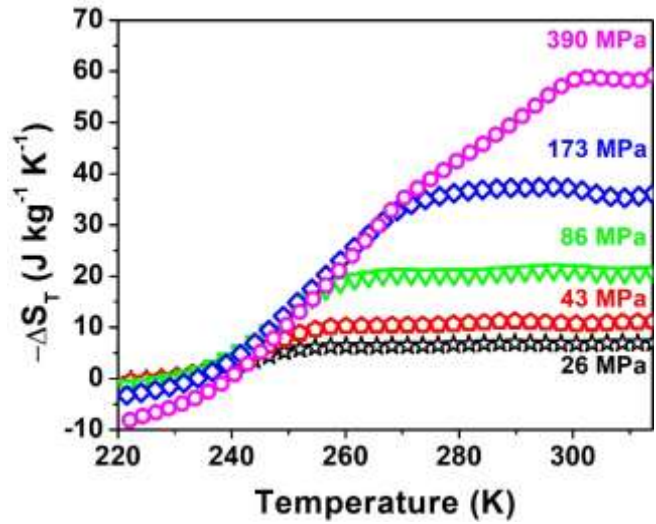

FIG. 3. Isothermal entropy change ($\Delta S_T$) vs temperature on heating process for NBR. Values for $\Delta\sigma$ = 26.0(5), 43(1), 86(2), 173(3) and 390(12) MPa, obtained from $\varepsilon$ vs T data at isobaric process (Fig. 2 and Fig. 3 in Supplemental Material [28]).

We have obtained giant values of $\Delta S_T$ for the heating process, as shown in Fig. 3. At 314 K and $\Delta\sigma$ = 390 MPa, a maximum $\Delta S_T$ of 59(6) J kg$^{-1}$ K$^{-1}$ was found. This leads to a normalized entropy change ($|\Delta S_T/\Delta\sigma|$) of 0.15(2) kJ kg$^{-1}$ K$^{-1}$ GPa$^{-1}$. A clear influence of $T_g$ is also observed in the profile of $\Delta S_T$ vs. T curve in Fig. 3: the plateau indicates the rubbery state, while the steep region is due to the glassy state.

At low temperatures (T $\leq$ 240 K) we observe positive $\Delta S_T$ values, trend associated with the inverse thermal expansion observed in the $\varepsilon$ vs. T data on heating (see Fig. 2 in Supplemental Material [28]), which is pronounced for the higher pressures. This inverse entropy change is probably related to the molecular rearrangements taking place in, since the movement of polymer chains slows down drastically in the glassy state. Therefore, the system requires a longer time-scale to reach the thermodynamic equilibrium when compared with the rubbery state [34]. As consequence of this mechanism, we suppose the structure is retained while the temperature cools down in such a way that, on heating, the structure is relaxed and contracts (inverse thermal expansion).

Envisaging the application of NBR in solid-state cooling, it is interesting to assess its performance under cyclic conditions. As we can see in ε vs. T data (see Fig. 4 in Supplemental Material [28]), NBR presents thermal hysteresis under cooling/heating cycles. Thermal hysteresis may be a drawback when considering the efficiency of refrigeration devices [41]. According to previous works [23,41], we calculated the reversible entropy changes ($\Delta S_{rev}$), defined as the intersection of the heating and cooling curves. This is represented by the shaded region in Fig. 4.

$\Delta S_{rev}$ fills a wide temperature range and slightly narrows as the pressure increases. Taking the 86 MPa curve [Fig. 4(a)], the temperature range goes from ~ 230 K to 314 K (range of 84 K). On the other hand, for 390 MPa [Fig. 4(c)], $\Delta S_{rev}$ was observed between 241 K and 314 K (range of 73 K), only 13% smaller than that for 86 MPa. Moreover, the losses in $\Delta S_T$ values in the reversible range are minimum due to the hysteresis and are mostly located at the left borders of the shaded region, especially considering the lower pressure variations (87 and 173 MPa). For the highest pressure change, (390 MPa), $\Delta S_{rev}$ is reduced by 15% (compared to the maximum $\Delta S_T$ in cooling process), but still maintaining its giant value of 56 J kg$^{-1}$ K$^{-1}$ [Fig. 4(c)].

Another interesting aspect of $\Delta S_{rev}$ region is its direct dependency on pressure. A good way to tackle this aspect is calculating the reversible refrigerant capacity ($RC_{rev}$) for several applied pressures. We can calculate $RC_{rev}$ by integrating $\Delta S_{rev}$ over the reversible temperature range. In Fig. 5(a), we observe a linear increment of $RC_{rev}$ values, which start to saturate at 234 MPa. The maximum $RC_{rev}$ of 2.7 kJ kg$^{-1}$ is achieved at 390 MPa, although at 234 MPa we already have a close value of 2.5 kJ kg$^{-1}$.

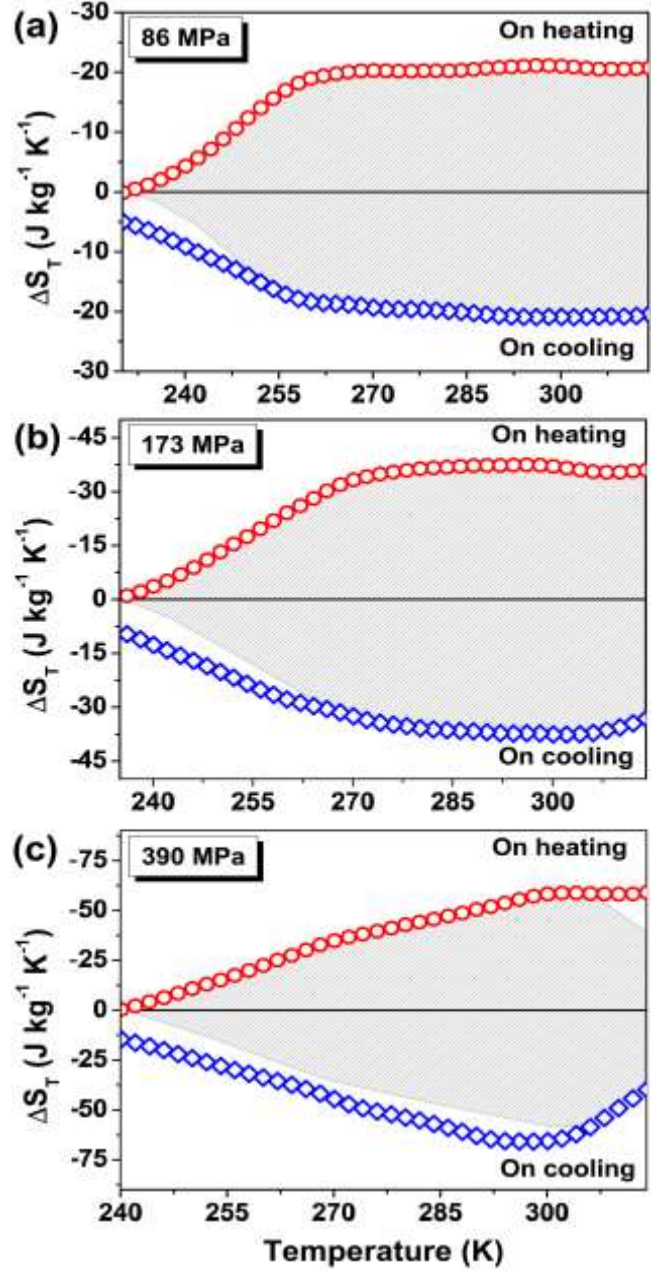

FIG. 4. Reversible entropy change for NBR. The red and blue curves are values for $\Delta S_T$ on heating and cooling processes, respectively. The shaded region corresponds to the reversible entropy change $\Delta S_{rev}$. Curves for pressures of 86 (a), 173 (b) and 390 MPa (c) are shown.

To compare the barocaloric performance of different materials, the normalized refrigerant capacity (NRC) is a useful parameter, which has been used before [3,4]:

$$NRC(\Delta T_{h-c}, \Delta\sigma) = \left| \frac{1}{\Delta\sigma} \int_{T_{cold}}^{T_{hot}} \Delta S_T(T, \Delta\sigma) dT \right|,$$

where $\Delta T_{h-c} \equiv T_{hot} - T_{cold}$ is the temperature difference between the hot ($T_{hot}$) and cold ($T_{cold}$) reservoirs, $\Delta\sigma$ is the pressure change. NRC as function of $\Delta T_{h-c}$ is shown in Fig. 5(b). For NBR, $T_{hot}$ was fixed at 315 K and $T_{cold}$ was varied between 310 K and 240 K for $\Delta\sigma = 173$

MPa. NRC values for NBR on the cooling process are similar to those obtained for heating process up to $\Delta T_{h-c}$ = 55 K. Above that point, the cooling process results in slightly higher values of NRC. Nevertheless, both heating and cooling present huge values of NRC, with a clear trend to increase.

When comparing NRC results of NBR with other materials, it is significantly higher than intermetallics $Mn_3GaN$ [42] and $Fe_{49}Rh_{51}$ [23], which both presenting a curve profile that tend to saturate at $\Delta T_{h-c}$ = 15 K. Comparing with PDMS, the values for NBR are higher for $\Delta T_{h-c}$ up to 20 K. Lastly, NBR presents lower NRC values when comparing with V-NR due to the higher $\Delta S_T$ values of the later material in the compared range. But for $\Delta T_{h-c}$ up to 10 K, both materials have similar values of NRC.

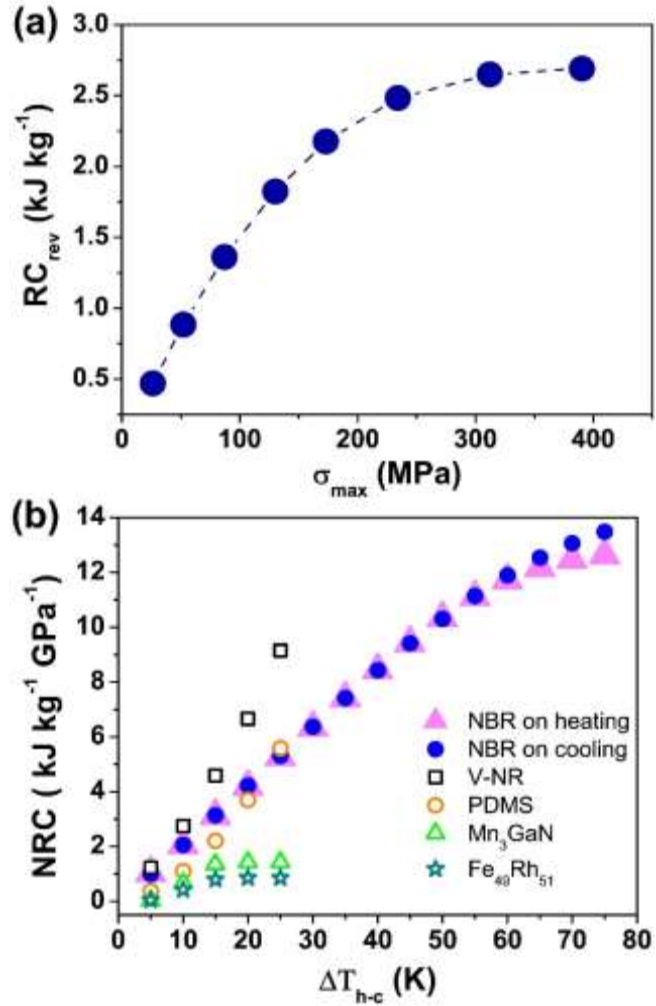

FIG. 5. (a) Reversible refrigerant capacity as a function of applied pressure. (b) Normalized refrigerant capacity as a function of $\Delta T_{h-c} = T_{hot} - T_{cold}$ (temperature difference between hot and cold reservoirs): NBR (both heating and cooling processes are shown; $T_{hot}$ = 315 K, $|\Delta\sigma|$ = 173 MPa); V-NR ( $T_{hot}$ = 315 K, $|\Delta\sigma|$ = 130 MPa); PDMS ( $T_{hot}$ = 315 K, $|\Delta\sigma|$ = 130 MPa) [3]; $Mn_3GaN$ ($T_{hot}$ = 295 K, $|\Delta\sigma|$ = 139 MPa) [42]; $Fe_{49}Rh_{51}$ ($T_{hot}$ = 325 K, $|\Delta\sigma|$ = 160 MPa) [23].

## IV. CONCLUSION

Our results on $\sigma_b$-CE of NBR demonstrate its promising potential regarding applications in solid-state cooling. Giant $\Delta T_S$ values of 16.4(2) K (at 323 K) and $\Delta S_T$ of 59(6)

J kg$^{-1}$ K$^{-1}$ (at 314 K) were obtained at relatively moderate applied pressures (390 MPa) and around room temperature. These results lead to huge normalized values: 0.15(2) J kg$^{-1}$K$^{-1}$GPa$^{-1}$ for |$\Delta S_T$/ $\Delta\sigma$| and 42(2) K GPa$^{-1}$ for |$\Delta T_S$/ $\Delta\sigma$|, comparable to the best barocaloric materials in the literature. Besides, $\Delta T_S$ demonstrated to be rather reversible: the compression and decompression results are very similar. Despite the hysteretic behavior of the material, reversible entropy change extends to a wide temperature range with minimal losses, keeping its giant values. NBR glass transition strongly influences $\sigma_b$-CE results, explaining the decreasing behavior of $\Delta T_S$ and $\Delta S_T$ values as the pressure change increases and temperature is reduced. Moreover, we determined the pressure coefficient of the glass transition in three distinct processes: isobaric strain vs. temperature curves on heating and on cooling processes, and from $\Delta T_S$ data.

## ACKNOWLEDGMENT

The authors acknowledge financial support from FAPESP (project number 2016/22934-3), CNPq, CAPES, LNLS and CNPEM.

## REFERENCES

[1] F. Edition, *The Science and Technology of Rubber* (2013).
[2] E. O. Usuda, N. M. Bom, and A. M. G. Carvalho, Eur. Polym. J. **92**, 287 (2017).
[3] N. M. Bom, W. Imamura, E. O. Usuda, L. S. Paixao, and A. M. G. Carvalho, ACS Macro Lett. **7**, 31 (2018).
[4] A. M. G. Carvalho, W. Imamura, E. O. Usuda, and N. M. Bom, Eur. Polym. J. **99**, 212 (2018).
[5] W. Imamura, E. O. Usuda, L. S. Paixão, N. M. Bom, and A. M. G. Carvalho, arXiv:1710.01761 (2017).
[6] S. Crossley, N. D. Mathur, and X. Moya, AIP Adv. **5**, 067153 (2015).
[7] L. Mañosa, A. Planes, and M. Acet, J. Mater. Chem. A **1**, 4925 (2013).
[8] I. Takeuchi and K. Sandeman, Phys. Today **68**, 48 (2015).
[9] V. K. Pecharsky and K. A. Gschneidner, Jr., Phys. Rev. Lett. **78**, 4494 (1997).
[10] A. S. Mischenko, Q. Zhang, J. F. Scott, R. W. Whatmore, and N. D. Mathur, Science **311**, 1270 (2006).
[11] L. Mañosa, D. González-alonso, A. Planes, E. Bonnot, M. Barrio, J. Tamarit, S. Aksoy, and M. Acet, Nat. Mater. **9**, 478 (2010).
[12] J. Gough, Memories Lit. Philos. Soc. Manchester **1**, 288 (1805).
[13] W. Thomson, Q. J. Math. **1**, 57 (1855).
[14] W. Thomson, Philos. Mag. Ser. **5**, 4 (1878).
[15] R. R. Wu, L. F. Bao, F. X. Hu, H. Wu, Q. Z. Huang, J. Wang, X. L. Dong, G. N. Li, J. R. Sun, F. R. Shen, T. Y. Zhao, X. Q. Zheng, L. C. Wang, Y. Liu, W. L. Zuo, Y. Y. Zhao, M. Zhang, X. C. Wang, C. Q. Jin, G. H. Rao, X. F. Han, and B. G. Shen, Sci. Rep. **5**, 1 (2015).
[16] M. Gorev, E. Bogdanov, I. Flerov, and N. Laptash, J. Phys. Condens. Matter **22**, 185901 (2010).
[17] M. V Gorev, E. V Bogdanov, I. N. Flerov, A. G. Kocharova, and N. M. Laptash, Phys. Solid State **52**, 167 (2010).
[18] M. V. Gorev, I. N. Flerov, E. V. Bogdanov, V. N. Voronov, and N. M. Laptash, Phys. Solid State **52**, 377 (2010).

[19] I. N. Flerov, A. V. Kartashev, M. V. Gorev, E. V. Bogdanov, S. V. Mel'Nikova, M. S. Molokeev, E. I. Pogoreltsev, and N. M. Laptash, J. Fluor. Chem. **183**, 1 (2016).
[20] C. Cazorla and D. Errandonea, Nano Lett. **16**, 3124 (2016).
[21] L. Mañosa, D. González-Alonso, A. Planes, M. Barrio, J. L. Tamarit, I. S. Titov, M. Acet, A. Bhattacharyya, and S. Majumdar, Nat. Commun. **2**, 1 (2011).
[22] S. Yuce, M. Barrio, B. Emre, E. Stern-Taulats, A. Planes, J. L. Tamarit, Y. Mudryk, K. A. Gschneidner, V. K. Pecharsky, and L. Mañosa, Appl. Phys. Lett. **101**, 071906 (2012).
[23] E. Stern-Taulats, A. Planes, P. Lloveras, M. Barrio, J. L. Tamarit, S. Pramanick, S. Majumdar, C. Frontera, and L. Mañosa, Phys. Rev. B **89**, 214105 (2014).
[24] P. Lloveras, E. Stern-Taulats, M. Barrio, J. L. Tamarit, S. Crossley, W. Li, V. Pomjakushin, A. Planes, L. Mañosa, N. D. Mathur, and X. Moya, Nat. Commun. **6**, 8801 (2015).
[25] E. Stern-Taulats, P. Lloveras, M. Barrio, E. Defay, M. Egilmez, A. Planes, J. L. Tamarit, L. Mañosa, N. D. Mathur, and X. Moya, APL Mater. **4**, 091102 (2016).
[26] J. M. Bermúdez-García, M. Sánchez-Andújar, S. Castro-García, J. López-Beceiro, R. Artiaga, and M. A. Señarís-Rodríguez, Nat. Commun. **8**, 15715 (2017).
[27] E. L. Rodriquez and F. E. Filisko, J. Appl. Phys. **53**, 6536 (1982).
[28] See Supplemental Material at [URL] for FT-IR and Strain-Temperature data for NBR sample
[29] N. M. Bom, E. O. Usuda, G. M. Guimarães, A. A. Coelho, and A. M. G. Carvalho, Rev. Sci. Instrum. **88**, 046103 (2017).
[30] L. Mañosa and A. Planes, Adv. Mater. **29**, 1603607 (2017).
[31] J. E. McKinney and M. Goldstein, J. Res. Natl. Bur. Stand. Sect. A Phys. Chem. **78A**, 331 (1974).
[32] J. J. Tribone, J. M. O'reilly, and J. Greener, J. Polym. Sci. Part B Polym. Phys. **27**, 837 (1989).
[33] H. A. Schneider, J. Therm. Anal. **47**, 453 (1996).
[34] J. M. Hutchinson, Prog. Polym. Sci. **20**, 703 (1995).
[35] T. S. Chow, *Mesoscopic Physics of Complex Materials* (Springer New York, New York, NY, 2000).
[36] C. A. Angell and W. Sichina, Ann. N. Y. Acad. Sci. **279**, 53 (1976).
[37] J. E. K. Schawe, Thermochim. Acta **603**, 128 (2015).
[38] T. V Tropin, J. W. Schmelzer, and V. L. Aksenov, Physics-Uspekhi **59**, 42 (2016).
[39] M. M. Vopson, J. Phys. D. Appl. Phys. **46**, 345304 (2013).
[40] Y. Liu, I. C. Infante, X. Lou, L. Bellaiche, J. F. Scott, and B. Dkhil, Adv. Mater. **26**, 6132 (2014).
[41] O. Gutfleisch, T. Gottschall, M. Fries, D. Benke, I. Radulov, K. P. Skokov, H. Wende, M. Gruner, M. Acet, P. Entel, and M. Farle, Philos. Trans. R. Soc. A Math. Phys. Eng. Sci. **374**, 20150308 (2016).
[42] D. Matsunami, A. Fujita, K. Takenaka, and M. Kano, Nat. Mater. **14**, 73 (2015).

# Supplemental Material:
# FT-IR and Strain-Temperature data for NBR sample

E. O. Usuda, [1,2] W. Imamura, [1, 3] N. M. Bom, [1] L. S. Paixão, [1] A. M. G. Carvalho, [1]

[1]*Laboratório Nacional de Luz Síncrotron (LNLS), Centro Nacional de Pesquisa em Energia e Materiais (CNPEM),CEP 13083-100, Campinas, SP, Brazil*

[2]*Departamento de Ciências Exatas e da Terra, Universidade Federal de São Paulo (UNIFESP), CEP 00972-270, Diadema, SP, Brazil.*

[3]*Faculdade de Engenharia Mecânica, Universidade Estadual de Campinas (UNICAMP), CEP 13083-860, Campinas, SP, Brazil.*

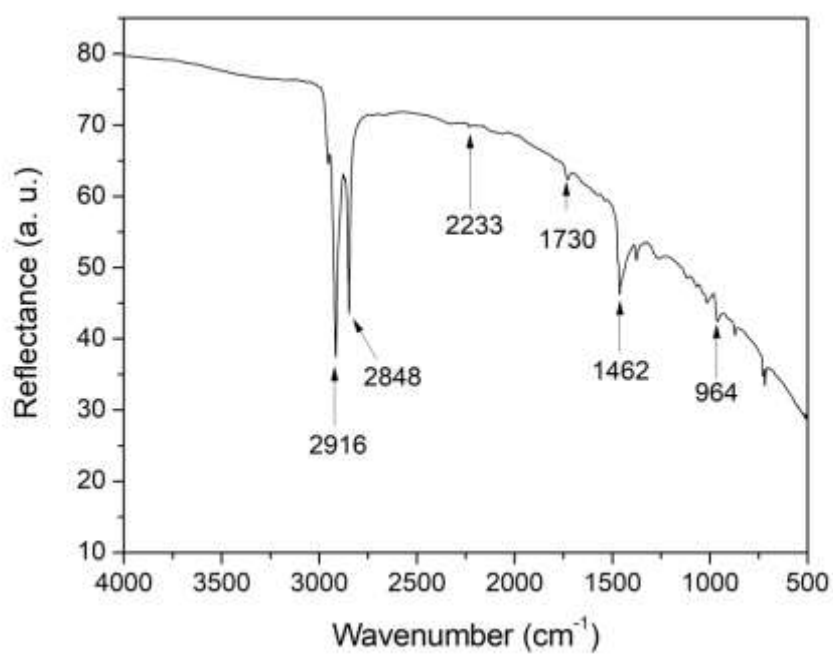

FIG. 1. FT-IR spectrum of NBR filled with carbon black. The bands are in accordance with the literature [1,2].

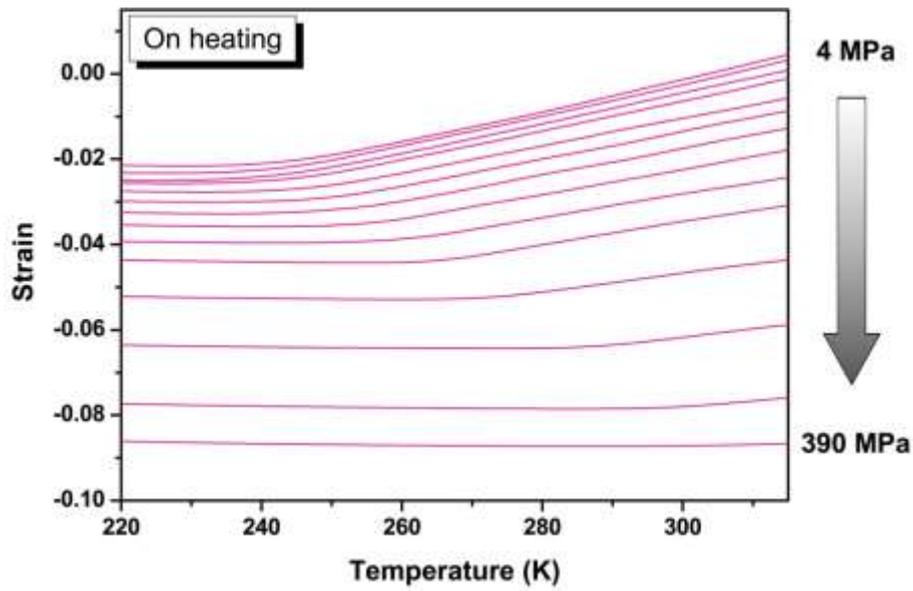

FIG. 2. Strain as a function of temperature for NBR on heating processes for various pressures (4 MPa – 390 MPa).

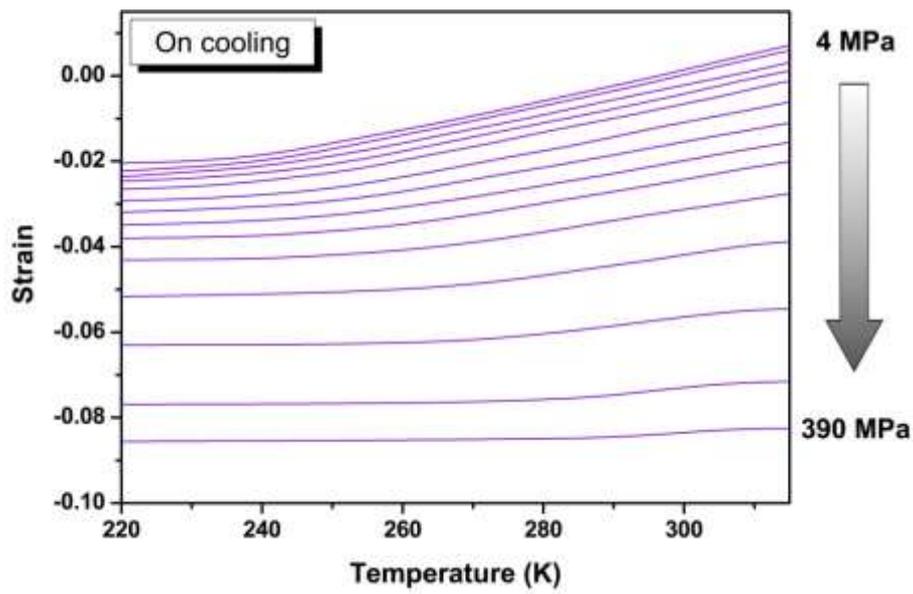

FIG. 3. Strain as a function of temperature for NBR on cooling processes for various pressures (4 MPa – 390 MPa).

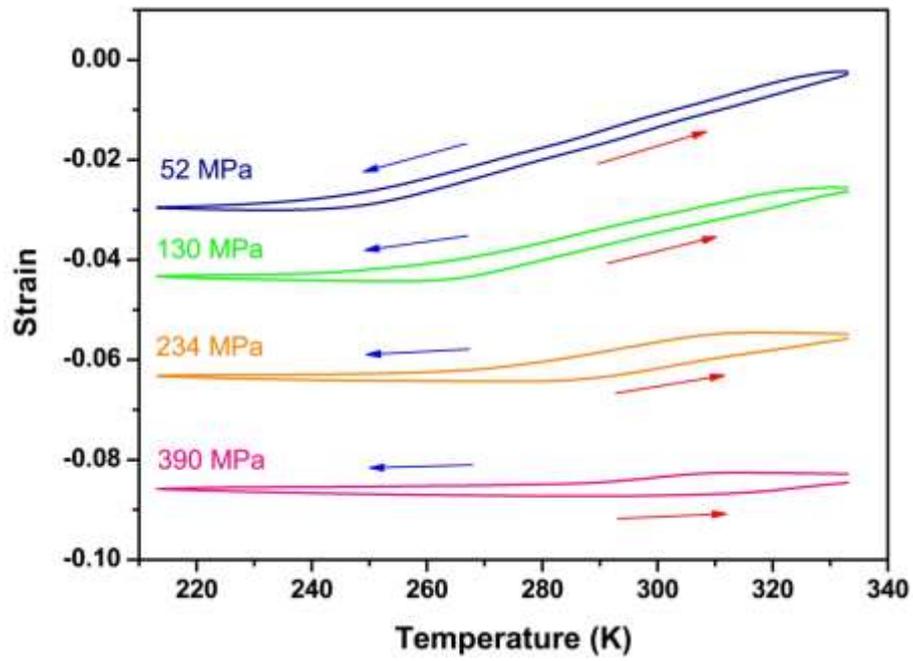

FIG. 4. Complete cycles of strain vs. temperature measurements for pressures of 52, 130, 234 and 390 MPa.

REFERENCES
[1] X. Liu, J. Zhao, R. Yang, R. Iervolino, and S. Barbera, Polym. Degrad. Stab. **151**, 136 (2018).
[2] J. Liu, X. Li, L. Xu, and P. Zhang, Polym. Test. **54**, 59 (2016).